\documentclass[a4paper]{article}
\usepackage{listings}
\usepackage{xcolor}
\usepackage{multirow}
\usepackage{jheppub}  
\usepackage{dcolumn}
\usepackage[english]{babel}
\usepackage[utf8x]{inputenc}
\usepackage[T1]{fontenc}
\usepackage{palatino}
\usepackage{amsmath}
\usepackage{afterpage}
\usepackage{graphicx, subcaption}
\usepackage[colorinlistoftodos]{todonotes}
\usepackage[colorlinks=true]{hyperref}
\usepackage{makeidx}
\newcommand{\be}{\begin{equation}}  
\newcommand{\ee}{\end{equation}}  
\newcommand{\bea}{\begin{eqnarray}}  
\newcommand{\eea}{\end{eqnarray}}  
\makeindex
\begin{document}

\vspace*{1.2cm}

\thispagestyle{empty}
\begin{center}
{\LARGE \bf $\rho$ photoproduction in ALICE}

\par\vspace*{7mm}\par

{\bigskip \large \bf Spencer R. Klein on behalf of the ALICE Collaboration}

\bigskip

{\large \bf  E-Mail: srklein@lbl.gov}

\bigskip

{Nuclear Science Division, Lawrence Berkeley National Laboratory, Berkeley CA 94720 USA}

\bigskip

{\it Presented at the Low-$x$ Workshop, Elba Island, Italy, September 27--October 1 2021}

\vspace*{15mm}

\end{center}
\vspace*{1mm}

\begin{abstract}

The $\rho^0$ is copiously photoproduced in ultra-peripheral heavy-ion collisions at the LHC.   In this talk, I will present recent results on rho photoproduction with ALICE, including cross-section measurements from PbPb and XeXe collisions, including a discussion of the production of high-mass final states that decay to $\pi^+\pi^-,$ and of neutron production that accompanies $\rho$ photoproduction.  I will conclude by presenting some prospects for ALICE in Runs 3 and 4. 

\end{abstract}
 
 \section{Introduction}
Ultra-peripheral collisions (UPCs) at heavy-ion colliders are a prolific source of photonuclear interactions; they are the energy frontier for photon-mediated reactions \cite{Baltz:2007kq,Bertulani:2005ru,Klein:2020fmr,Contreras:2015dqa}.   In UPCs, the ions do not interact hadronically, so the product of the photon-mediated interaction is visible.  In a simple approximation, the impact parameter $b$ must be more than twice the nuclear radius, $R_A$. UPCs include both two-photon interactions and photoproduction.   UPCs at the Large Hadron Collider (LHC) represent the energy frontier for photon-mediated interactions.  In pp collisions, photon-proton center of mass energies up to about 3 TeV are accessible, while in pPb collisions, the maximum energy is about 700 GeV. The photons have a  small $p_T$, roughly $p_Z/\gamma$, where $\gamma$ is the ion Lorentz boost, so it is possible to use the $p_T$ distribution to probe the size of the nuclei.  Unfortunately, the photon $p_T$ is a conjugate variable to the impact parameter, so if restrictions are imposed on $b$ (such as $b>2R_A$), the mean $p_T$ will increase; this increase is not calculable without some assumptions \cite{Klein:2020jom}. 

In vector meson photoproduction, an incident photon fluctuates to a quark-antiquark dipole which then scatters elastically from a target nucleus, emerging as a real vector meson.   During the elastic scattering, the vector meson retains the same quantum numbers (including helicity) as the incident photon.

The $\rho^0$ is of special interest as the most copiously photoproduced vector meson \cite{STAR:2002caw}.  It is the lightest vector meson, corresponding to the largest dipole, so is the most subject to nuclear effects.    In addition to $\rho$ photoproduction, the photon can fluctuate directly to a $\pi^+\pi^-$ pair, which then scatters, emerging as a real pion pair \cite{Bauer:1977iq}.    These two possibilities are indistinguishable, so interfere with each other, enhancing the spectrum below the $\rho^0$ mass, and suppressing it at higher masses. Higher-mass, excited $\rho$ states are also possible, and can lead to higher mass $\pi^+\pi^-$ pairs. 

One complication in PbPb UPCs is that the coupling constant $Z\alpha\approx 0.6$ is large, so for a collision with moderate impact parameters, it is very possible to exchange more than one photon, complicating the reaction \cite{Baltz:2002pp,Baur:2003ar}.   A second photon is likely to excite one of the nuclei, while a third photon may excite the other nucleus, leading to mutual Coulomb dissociation \cite{Baltz:1996as}.  The excitation may be collective, such as a Giant Dipole Resonance (GDR) or higher excitation, or an excitation of a single nucleon to a $\Delta$ or higher resonance, or, for higher energy photons, a more complex hadronic interaction.  Production of an additional vector meson is also possible.   Most of these reactions involve nuclear dissociation, leading to the emission of one or more neutrons, or, less frequently, one or more protons.   It is also possible to produce a vector meson and excite the nucleus via one-photon exchange leading to incoherent photoproduction.  

 In these reactions, the photons are emitted independently  \cite{Baur:2003ar}, connected only by a common impact parameter.  The impact parameter affects the photon flux and maximum energy.  And, since the photons are polarized with their electric field vectors parallel to the impact parameter vector, the photons share the same polarization.

These additional reaction products complicate the analysis of UPC photoproduction, since one can no longer focus exclusively on exclusive reactions.   In most cases, the additional reaction only leads to the production of neutrons, but sometimes $\pi^{\pm}$ may be created.      The cross-section for having multiple reactions may be computed in impact parameter ($b$) space.  For example, the cross-section to produce a $\rho$ with multiple Coulomb excitation is
\begin{equation}
\sigma = \int {\rm d}^2b P_{\rho} (b) P_{X1}(b) P_{X2}(b)
\label{eq:mult}
\end{equation}
where $P_{\rho} (b)$,  $P_{X1}(b)$, and  $P_{X2}(b)$ are the probabilities to produce a $\rho$, and excite the first and second nuclei respectively.   These probabilities are given by the product of the differential photon flux ${\rm d}^2N_\gamma/{\rm d}b^2$ and the $\gamma A$ cross-sections.   For nuclear excitations, the $\gamma A$ cross-sections must include a wide range of reactions, including collective nuclear excitations like the Giant Dipole Resonance (GDR), nucleon excitations, such as the $\Delta^+$ resonance, and partonic excitations from high-energy photons, spanning a wide photon energy range from about 10 MeV (in the target frame) up to the kinematic limit.   These cross-sections are usually determined using tabulations of data from multiple sources \cite{Baltz:1996as}.   When the cross-sections are large, it may be necessary to include a unitarity correction, since multiple photons may contribute to excitate a single target to a higher level.

For photons below a cutoff energy (when $k< \gamma \hbar c/b$), the photon flux has a $1/b^2$ dependence, so, the more photons that are exchanged, the smaller  $\langle b\rangle$  \cite{Baur:2003ar}.  So, one can use the number of exchanged photons to preferentially select different ranges of impact parameter. They also all share the same photon polarization, so polarization correlations are expected. 

\section{Detector and Data Analysis}

The data used here were collected using the ALICE detector, which comprises a large central detector and a forward muon spectrometer \cite{ALICE:2008ngc}.  For the analyses discussed here, the most important components are an inner silicon detector and large time projection chamber, in a 0.5 T solenoidal magnet. 

The events analyzed here were collected with a special trigger optimized for ultra-peripheral collisions \cite{ALICE:2020ugp,ALICE:2021jnv}.  It required two pairs of signals in the silicon detector, with azimuthal angular separation greater than 153 degrees.  Each signal pair, consisting of hits in different silicon layers, was consistent with one track.  The azimuthal angle requirement was to select pairs where the tracks were roughly back-to-back.  

The other trigger requirements provided vetos to reject events that contained additional particles.  The four veto detectors, and their pseudorapidity ($\eta$) coverage are: V0A ($2.8 < \eta < 5.1$), V0C ($-3.7 < \eta<-1.7$), ADA ($4.7 < \eta < 6.3$) and ADC ($-4.9 > \eta > -6.9$). 

Data from the zero degree calorimeters (ZDCs) were used in the analysis, but not in the trigger.

The analysis selected events with exactly two good oppositely charged tracks, consistent with a vertex in the interaction region.  The tracks were required to have specific energy loss (${\rm d}E/{\rm d}x$) in the TPC consistent with being $\pi^\pm$.

\begin{figure}
\begin{center}
\epsfig{figure=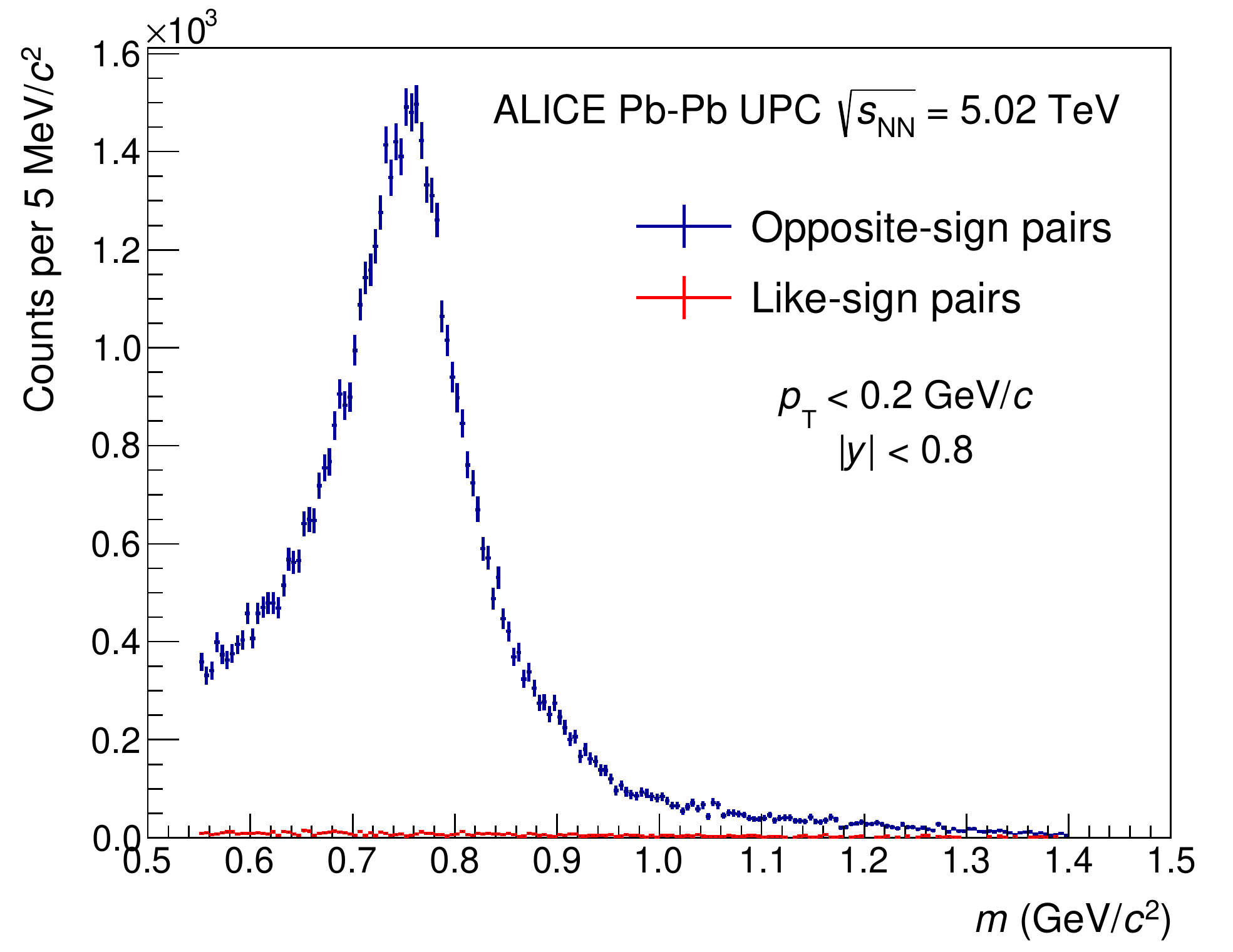,height=0.35\textwidth}
\epsfig{figure=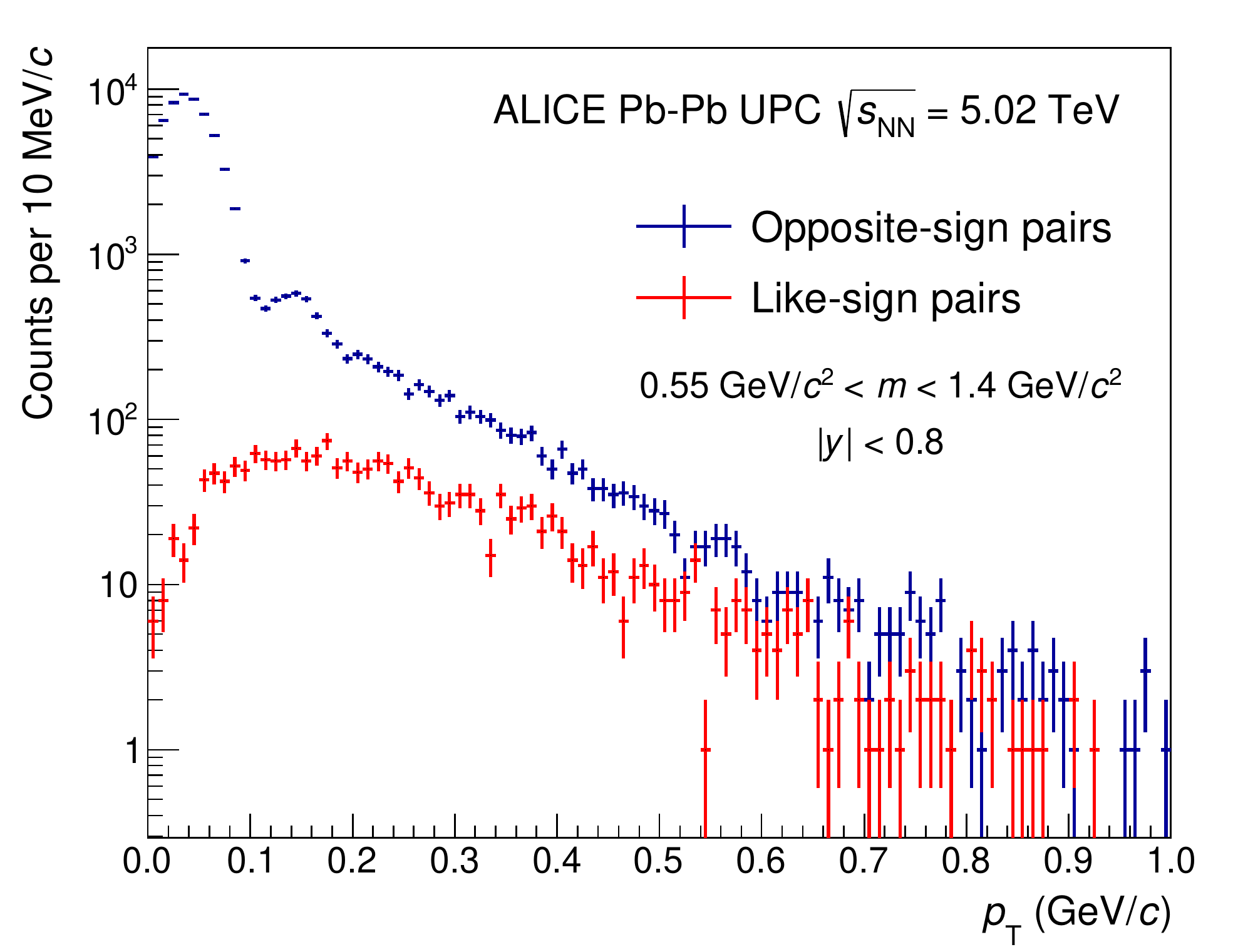,height=0.35\textwidth}
\caption{The mass (left) and $p_T$ spectra for selected pairs.  A cut of $p_T<200$ MeV/c is applied to the mass spectrum, to emphasize coherent production, while the $p_T$ spectrum includes a wide range of masses.  For $p_T < 200$ MeV/c, the like-sign background is orders of magnitude below the coherent production signal.  From \cite{ALICE:2020ugp}.}
\label{fig:first}
\end{center}
\end{figure}

\section{$\rho$ production results}

These cuts left a clean signal.  Figure \ref{fig:first} shows the dipion invariant mass and $p_T$ spectra.  The two peaks in the $p_T$ spectrum below  200 MeV/c, correspond to the first two diffractive maxima, clearly showing the diffractive nature of the production.  The like-sign pairs, which are a proxy for most backgrounds (notably grazing hadronic collisions), are far below the oppositely charged pairs, showing that there is little background in the events.  Two remaining backgrounds, from photoproduction of the $\omega$, followed by $\omega\rightarrow\pi^+\pi^-\pi^0$ and $\gamma\gamma\rightarrow\mu^+\mu^-$ are also small, and mostly concentrated at low $M_{\pi\pi}$.  The latter is a background since we cannot effectively distinguish $\mu^\pm$ and $\pi^\pm$. 

The $M_{\pi\pi}$ mass spectrum can be fit with a relativistic Breit-Wigner distribution for the $\rho^0$ plus a constant term for direct $\pi^+\pi^-$, and an interference term between the two.  $\gamma\gamma\rightarrow\mu^+\mu^-$ is included with a template based on STARlight \cite{Klein:2016yzr}, while $\omega\rightarrow\pi^+\pi^-\pi^0$ is removed by only fitting in the region $M_{\pi\pi}>0.62$ GeV.

\begin{figure}
\begin{center}
\epsfig{figure=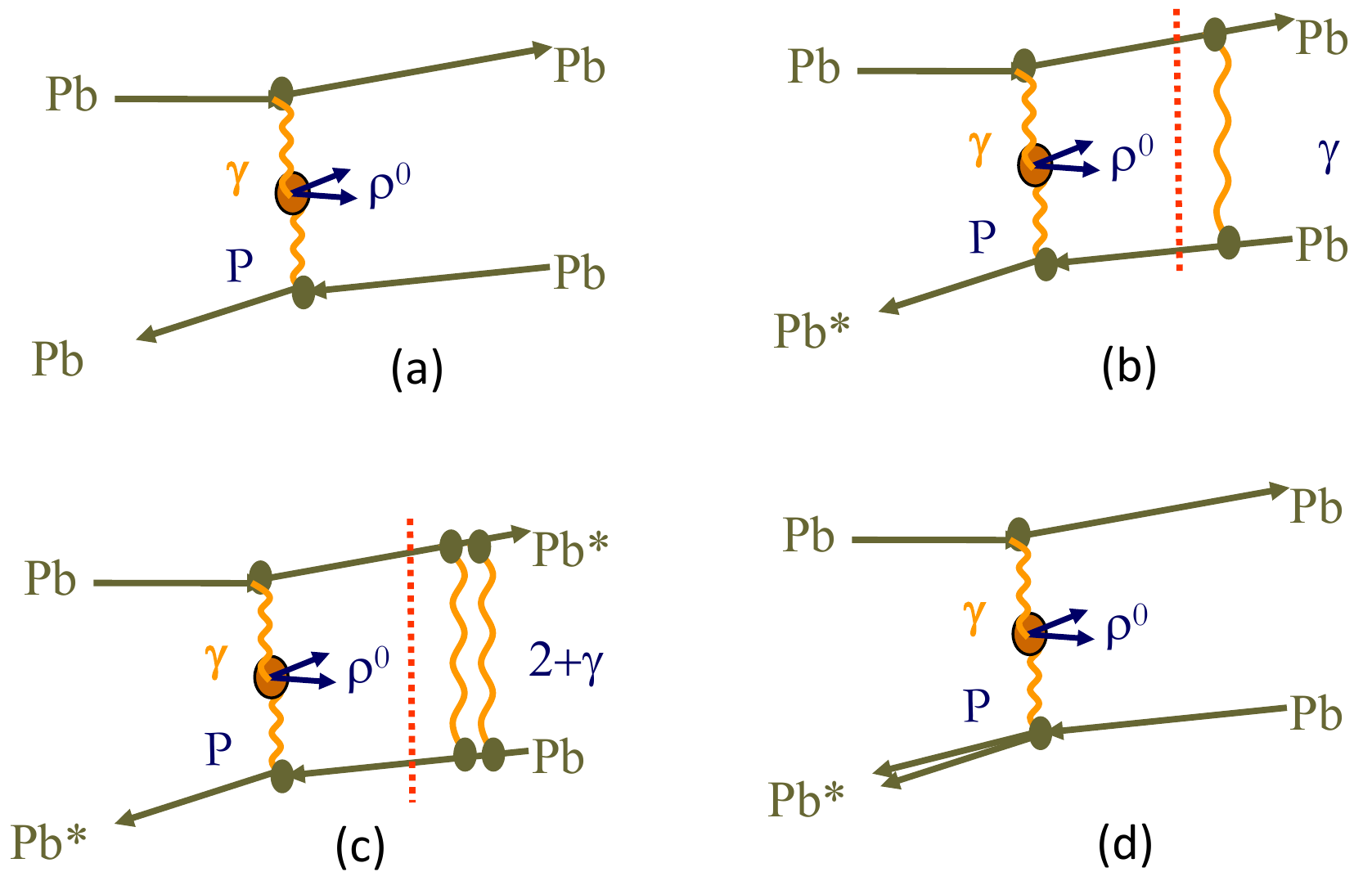,height=0.3\textwidth}
\epsfig{figure=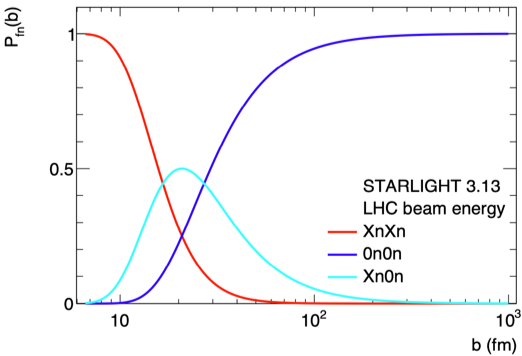,height=0.3\textwidth}
\caption{(left) Four diagrams that contribute to $\rho$ photoproduction: (a) coherent production, plus coherent production with an additional (b) one  or (c) two photons exchanged, and (d) incoherent photoproduction.  The products from reactions (b) and (d) are not completely distinguishable.   (Right) The impact parameter distributions for different nuclear excitations: no nuclear excitation (0n0n, diagram (a), single nuclear excitation (0nXn, diagram (b)) and mutual Coulomb excitation (XnXn, diagram (c)).  Nuclear excitation preferentially selects events with smaller impact parameters.  Incoherent photoproduction corresponds to single-photon exchange, so it has a similar impact parameter distribution as 0n0n.  The right panel is from Ref. \cite{Klein:2020fmr}.
}
\label{fig:diagrams}
\end{center}
\end{figure}

The produced $\rho$ and direct $\pi^+\pi^-$ may be accompanied by neutrons, which can occur when the two nuclei exchange additional photons, as in Figs. \ref{fig:diagrams}(b) and (c), or from 
an incoherent photoproduction reaction (Figs. \ref{fig:diagrams}(d)) involving a single photon.    The photons are expected to be emitted independently, sharing only a common impact parameter \cite{Gupta:1955zza,Baur:2003ar}.      

\begin{figure}
\begin{center}
\epsfig{figure=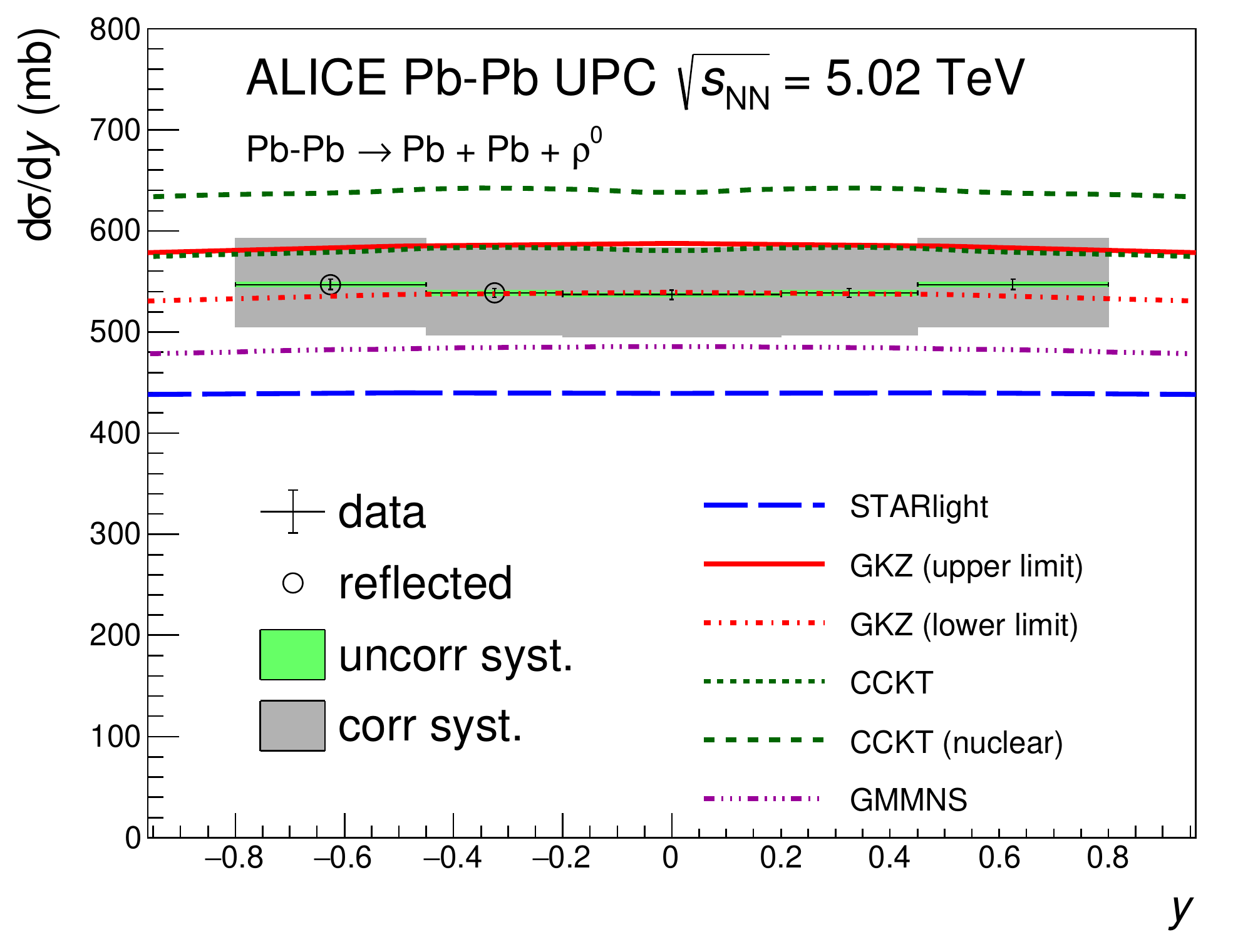,height=0.35\textwidth}
\epsfig{figure=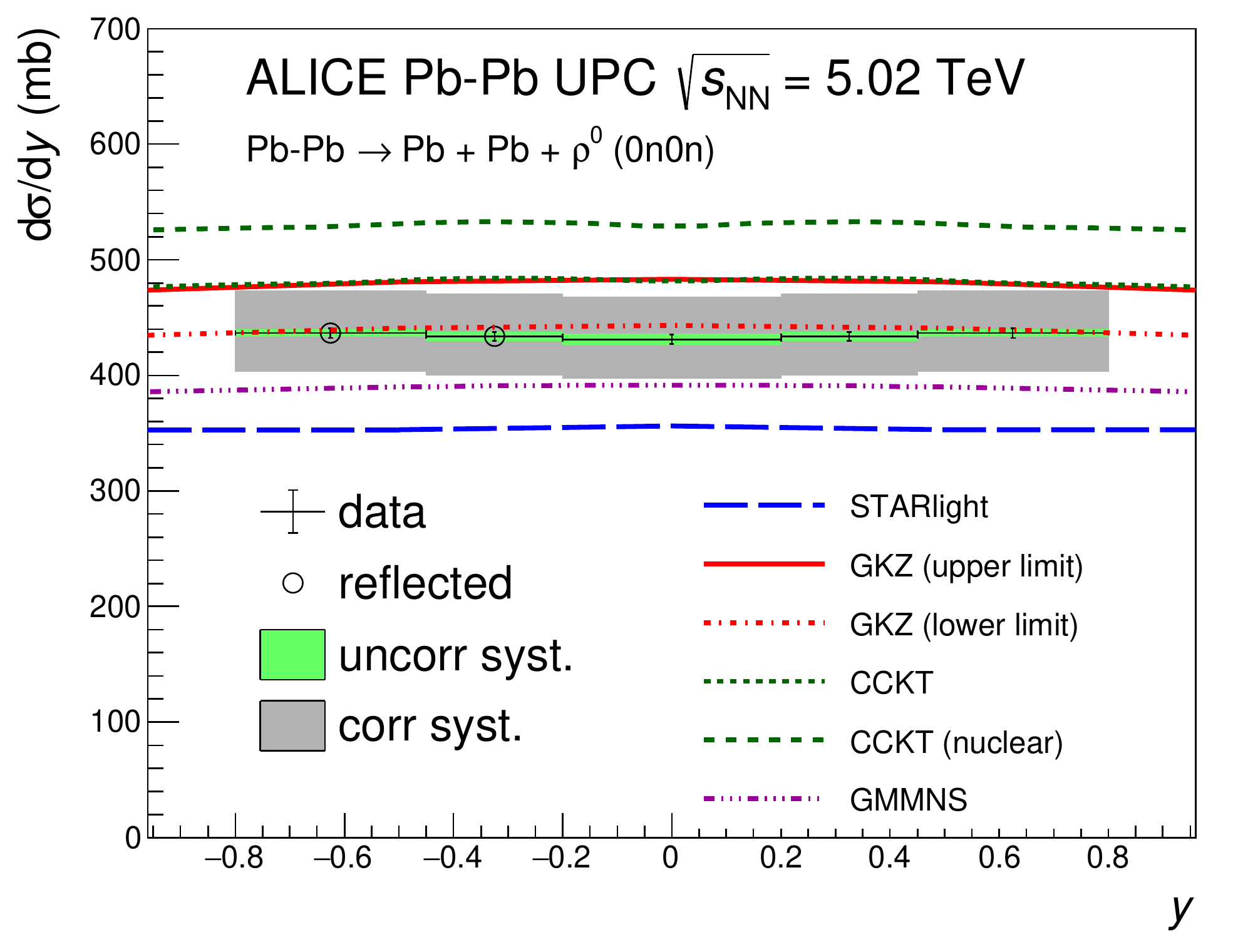,height=0.35\textwidth}
\epsfig{figure=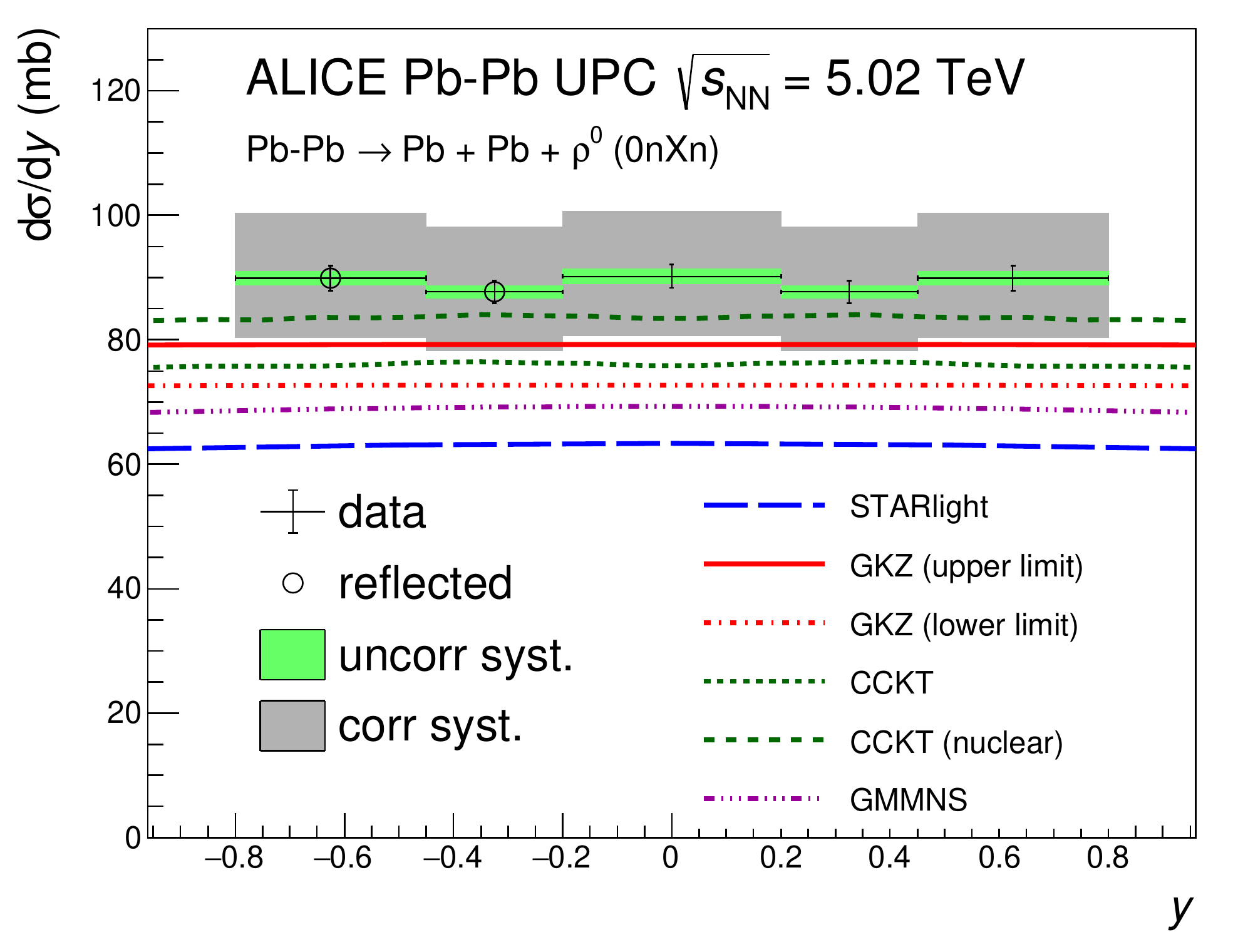,height=0.35\textwidth}
\epsfig{figure=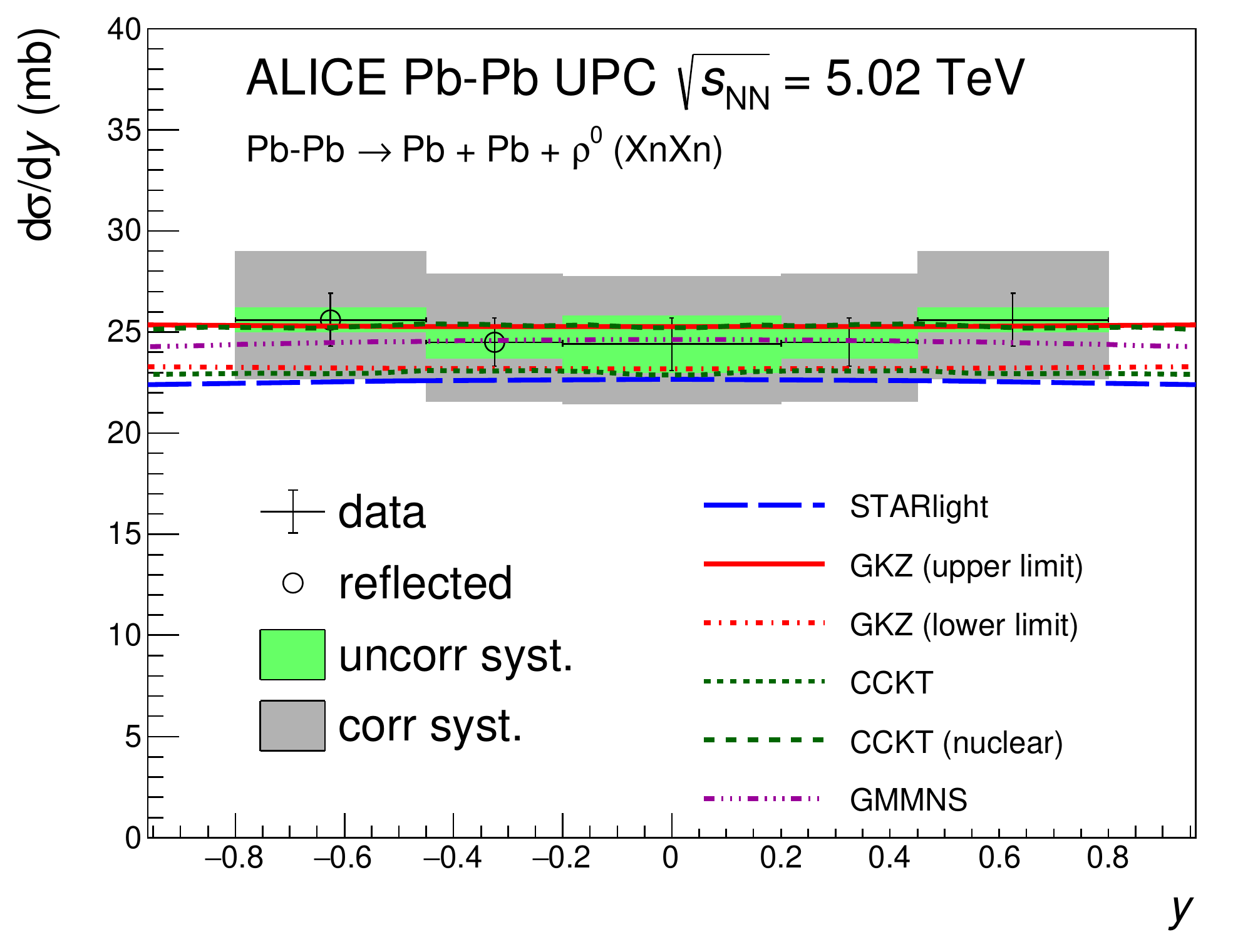,height=0.35\textwidth}
\caption{$d\sigma/dy$ for $\rho$ photoproduction for (top left) all events, and three different classes of neutron emission: (top right) no neutrons, (bottom left) neutrons in one ZDC only, and (bottom right) neutrons in both ZDCs.  Each panel is compared with several different theoretical calculations.  From
\cite{ALICE:2020ugp}.
}
\label{fig:dsdt}
\end{center}
\end{figure}

Figure \ref{fig:dsdt} shows the cross-sections for all events, and for the same events divided into three classes.   The top-left panel shows the total measured $\rho$ cross-section, compared to five models.  STARlight is based on parameterized HERA $\gamma {\rm p}$ data, with a Glauber-like eikonal formalism to handle nuclear targets \cite{Klein:1999qj}.   The GKZ predictions are based on a modified vector-meson-dominance model, using a Glauber-Gribov formalism for nuclear targets \cite{Guzey:2013jaa}.  The Glauber-Gribov approach allows for a dipole to interact multiple times as it traverses a target.  Each individual interaction can be inelastic, with the intermediate states (between interactions) allowed to include high-mass fluctuations. The CCKT predictions are based on a calculation of dipoles passing through a nuclear target, which is modeled in terms of gluon density as a function of impact parameter \cite{Cepila:2016uku}.  The gluon density includes gluonic hot-spots.  Finally, the GMMNS model is another dipole based calculation that includes an implementation of gluon saturation \cite{Goncalves:2017wgg}.   Most of the models do a reasonable job of matching the data, although STARlight is a bit on the low side, and the CCKT (nuclear) model is a bit high.

Per Eq. \ref{eq:mult}, the cross-sections for additional photon exchange may be easily calculated given a $\sigma_{\gamma p}$.  
The STARlight neutron calculation is done within the STARlight code \cite{Klein:2016yzr}, while the CCKT simulation used the {\bf $n_0^0n$} afterburner \cite{Broz:2019kpl}.  To the extent that these calculations are based on the same parameterized photoexcitation data, they should give the same relative cross-sections for $\rho^0$ production accompanied by neutron emission.   However, the relative cross-sections do not agree perfectly.  For example, in the two upper panels (total coherent $\rho$ cross-section and $\rho$ without neutron emission), 
the CCKT (nuclear) cross-section is well above the other calculations, while in the lower panels, where neutron emission is required, it is relatively lower.  A similar trend is evident for the CCKT curve.   It may be that {\bf $n_0^0n$} predicts lower excitation probabilities than STARlight. 

Experimentally, neutron emission, expected in most photonuclear reactions, is easy to detect using the ALICE ZDCs.  However, there is a complication.  Some of the photoexcitation occurs at high energies, and the reactions can lead to emission of one or more $\pi^\pm$ or heavier particles.  If these particles hit any of the detectors used as event vetoes (the ADA, ADC, V0A and V0C), causing a loss of $\rho$ signal.    The loss is substantial; it is $26\pm4\%$ for events with neutrons in one ZDC, and $43\pm5\%$ for events with neutrons in both ZDCs.   This loss is estimated using control triggers which do not include the veto, and appropriate corrections are applied.  
 
\section{$\rho$ photoproduction in XeXe collisions}

The $A$ dependence of $\rho$ photoproduction can provide an important clue about the presence of saturation or other high-density nuclear phenomena.     In 2017, the LHC collided xenon atoms, at $\sqrt{s_{NN}}=5.44$ TeV.  ALICE used the same UPC trigger as for lead-lead running and measured the cross-section, using similar methods \cite{ALICE:2021jnv}.  Figure \ref{fig:highmass} shows the $\rho$ photoproduction as a function of atomic number.   At mid-rapidity,
\begin{equation}
\frac{{\rm d}\sigma}{{\rm d}y} = 131.5 \pm 5.6 ({\rm stat.})^{+17.5}_{-16.9} ({\rm syst.})\ {\rm mb}.
\end{equation}
This is slightly below the STARlight predictions, slightly below the lower bound of the GMMNS prediction, and below the GKZ band.  However, none of these deviations are very significant. 

The cross-section scales as $A^\alpha$, with $\alpha=0.96\pm0.02$, dominated by systematic uncertainty.  This shows the presence of substantial nuclear effects.  Without nuclear effects, the coherent cross-section would scale as $A^{4/3}$.  This is the product of two scaling relations: the forward cross-section scales as $A^2$, while the $p_T$ range over which coherent production is possible scales as $A^{-2/3}$, leaving the $4/3$ exponent.   On the other hand, it is also considerably above the prediction of a black disk model, in which the cross-section scales as the frontal area of the nucleus, $A^{2/3}$.  

\section{A high-mass state}

\begin{figure}
\begin{center}
\epsfig{figure=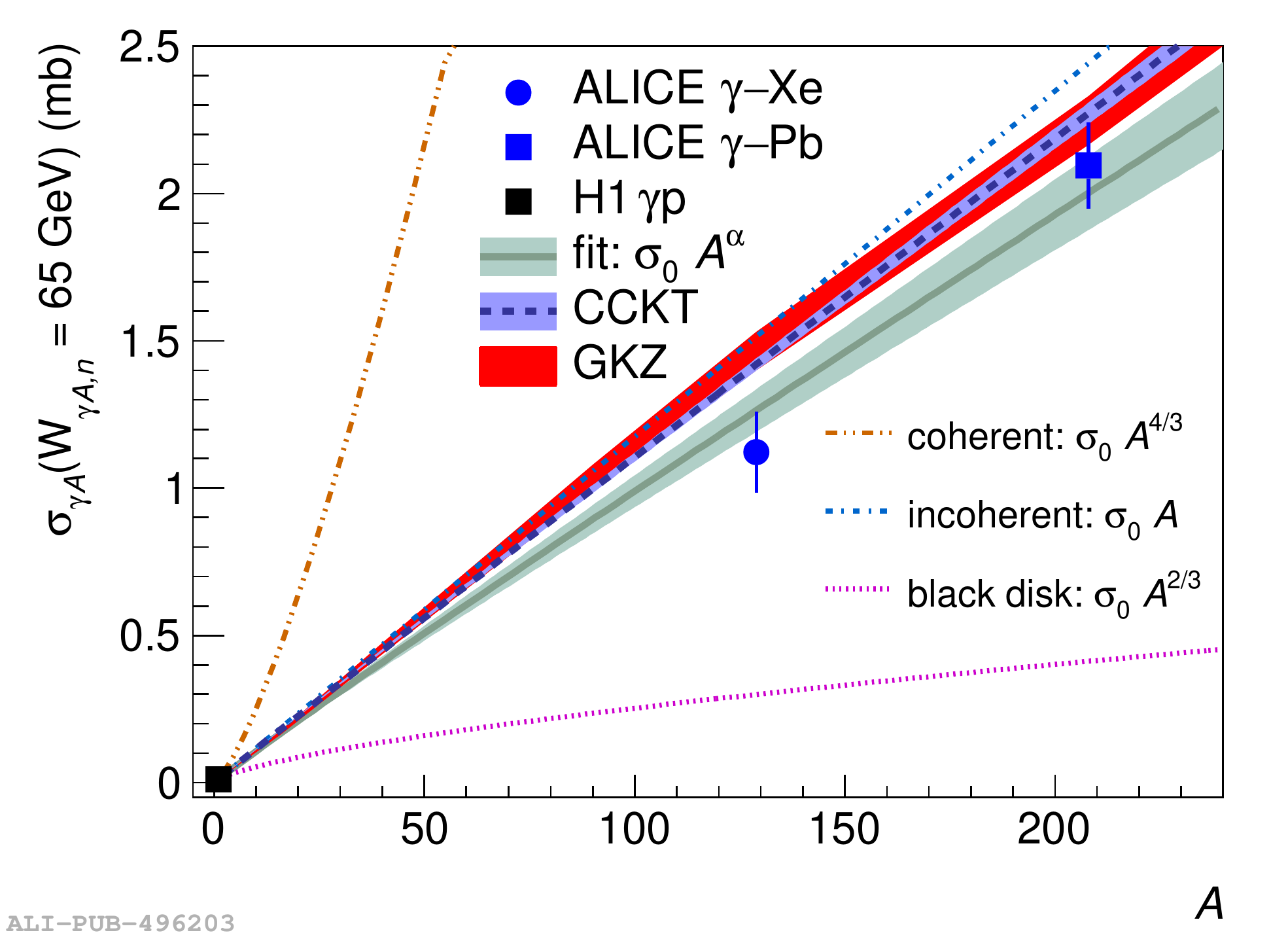,height=0.3\textwidth}
\epsfig{figure=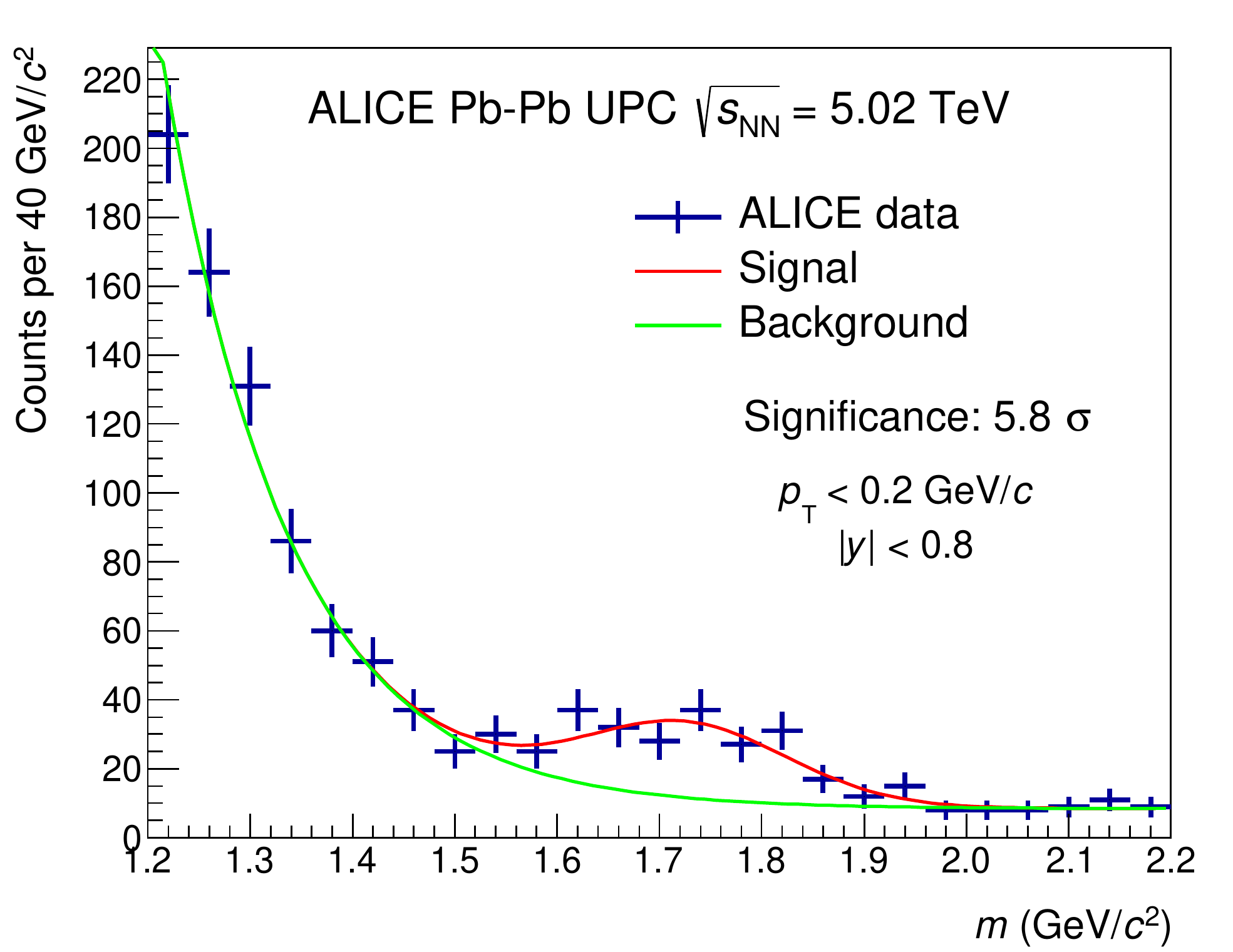,height=0.3\textwidth}
\caption{(left) $\gamma A\rightarrow \rho A$ cross-sections vs. atomic number $A$ for pp, XeXe and PbPb collisions, for 65 GeV photons. (right)$M_{\pi\pi}$ for exclusive production for events with pair $p_T<200$ MeV/c.   A broad resonance is visible over the high-mass tail of the $\rho^0$ plus direct $\pi^+\pi^-$.  From \cite{ALICE:2020ugp}.
}
\label{fig:highmass}
\end{center}
\end{figure}

A number of excited, higher-mass $\rho$ states can decay to $\pi^+\pi^-$.  Figure \ref{fig:highmass} (right) shows the $\pi^+\pi^-$ mass spectrum for events with $p_T < 200$ MeV/c in lead-lead collisions.  The expected tail of the $\rho^0$ is visible, with a broad resonance on top of it.   The spectrum is fit using an exponential for the  $\rho^0$ plus the direct $\pi^+\pi^-$ tail, plus a Gaussian for the resonance.  The null (no-resonance) hypothesis is rejected with 4.5 $\sigma$ significance.  The resonance best-fit parameters are mass $M=1725\pm 17$ MeV and width $\Gamma=143\pm21$ MeV.    The resonance is similar to that seen by STAR in gold-gold UPCs, with $M=1653\pm 10$ MeV and width $\Gamma=164\pm15$ MeV \cite{Klein:2016dtn}.   STAR pointed out that the peak might be compatible with photoproduction of the $\rho_3 (1690)$.  The ZEUS Collaboration also saw resonances in exclusive $\pi^+\pi^-$ electroproduction ($Q^2>2$ GeV$^2$), with masses of $1350 \pm 20  ^{+20}_{-30}$ MeV and $1780 \pm 20 ^{+15}_{-20}$ MeV \cite{ZEUS:2011tzw}.

\section{Future plans}

During LHC Run 3 and Run 4, ALICE will have many improvements which will improve charged particle reconstruction, raise ALICE's rate capability and remove trigger bottlenecks for UPC data collection.   The TPC endcaps have been replaced with GEM based readouts to allow continuous (rather than gated) TPC readout and a new ITS2 silicon tracker will use monolithic active pixel sensors to greatly improve vertexing, especially for open charm hadrons.   

For UPCs, the biggest improvement will be a new streaming readout which will do away with triggering.  All data will flow to the data acquisition system, where it can be studied with high-level event selection algorithms \cite{Antonioli:2013ppp}; for lead-lead running, all data will be saved.   This will give an enormous boost to UPC data collection, since triggering is usually the limiting factor for UPC studies.  During  Run 3 and Run 4, a total of 13 pb$^{-1}$ of lead-lead data should be collected.  This is equivalent to 5.5 billion $\rho^0\rightarrow\pi^+\pi^-$ within the ALICE acceptance, along with 210 million $\rho'\rightarrow\pi^+\pi^-\pi^+\pi^-$.  The $J/\psi$ sample should include 1.1 million $J/\psi\rightarrow\mu^+\mu^-$ in the central detector, a similar number of $e^+e^-$ plus about 600,000 $\mu^+\mu^-$ in the forward spectrometer \cite{Citron:2018lsq}.  For the $\psi'$, the rates are about 35,000 and 19,000 respectively. Photoproduction of $\Upsilon(1S)\rightarrow\mu^+\mu^-$ should also be visible, with 2,800 events expected in the central detector, and 880 in the forward muon spectrometer.   This should be enough for detailed studies of the spectroscopy of the light vector mesons, including of the substructure of the heavier mesons. It should also be possible to measure the production characteristics of heavy quarkonium, comparing the effect of shadowing on mesons of different masses.  The removal of the trigger bias and the improved vertex measurements will also facilitate the study of photoproduction of open charm.  

In addition to lead-lead running a short (0.5 nb$^{-1}$?) oxygen-oxygen run has been proposed for Run 3 \cite{ALICE:2021wim,Brewer:2021kiv}.  This offers two unique opportunities for UPCs. 

The first is to study incoherent photoproduction of the $\rho$ on oxygen targets.   Incoherent photoproduction is of great interest because, in the Good-Walker paradigm, it is related to event-by-event fluctuations in the nuclear configuration - the phase space that includes both the individual nucleon positions and, more importantly, the presence of gluonic `hot spot' fluctuations \cite{Klein:2019qfb}.  The $p_T$ spectrum for incoherent production is loosely tied to the length scale for these fluctuations, so it is desirable to measure over as wide a range in $p_T$ as possible.
  
\begin{figure}
\begin{center}
\epsfig{figure=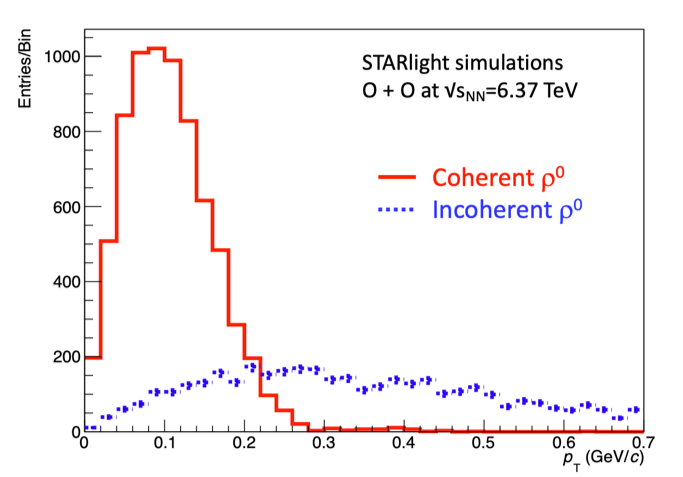,height=0.3\textwidth}
\caption{(left) Simulated $p_T$ spectrum for coherent and incoherent $\rho$ photoproduction in oxygen-oxygen collisions at $\sqrt{s_{NN}}=6.37$ TeV.   From \cite{ALICE:2021wim}.
}
\label{fig:oxygen}
\end{center}
\end{figure}

It is difficult to study incoherent photoproduction in lead-lead collisions because of the large background from coherent production.   In oxygen-oxygen running, the ratio of incoherent to coherent production is expected to be larger than in lead-lead.  As Fig. \ref{fig:oxygen} shows, the predicted coherent peak is still larger than the incoherent, but by less than in lead-lead collisions.  The other difference with lead-lead collisions is that oxygen is only charge eight, so most reactions only involve single photon exchange. Multi-photon exchange, as discussed above, is almost absent, so the presence of neutrons is a clear sign of nuclear breakup.  Although not all nuclear dissociation will produce neutrons, most will do so.   Photonic excitation is also possible, but oxygen is a very stable, doubly magic nucleus, with a lowest lying excited state at 6.05 MeV \cite{IAEA}, so nucleon emission is likely to be predominant. 

The second opportunity is to study the competition between photoproduction and double-diffractive interactions.  Both of these reactions can lead to $\pi^+\pi^-$ final states.  Photoproduction dominates in heavy-ion collisions, while double-diffractive interactions dominate in pp collisions.  With medium-charge nuclei, the amplitudes should be similar, so interference may be possible among final states with the same spin/parity.  

\section{Conclusions}

The $\rho^0$ is copiously photoproduced in ultra-peripheral collisions of heavy ions.  The cross-section for coherent $\rho$ photoproduction is quite well reproduced in models that use Glauber or dipole calculations to predict the cross-sections.   The cross-section scales with the atomic number $A$ as $A^{0.96\pm 0.02}$,  showing that nuclear effects substantially moderate the $A^{4/3}$ dependence expected for full coherence, without nuclear suppression.

The cross-section for coherent $\rho$ photoproduction accompanied by neutron emission is consistent with a model whereby the neutron production comes through the exchange of one or more additional photons, which are independent of the $\rho$ production, except for sharing a common impact parameter.  

We have also observed a heavy state, with a mass of 1650 MeV, decaying to $\pi^+\pi^-$.  The mass and cross-section may be consistent with that expected for the $\rho_3(1690)$. 

Looking ahead, ALICE expects a rich bounty of UPC results during LHC Run 3 and Run 4.   The new flow-through data acquisition system will eliminate the bottleneck formerly imposed by the requirements of a low-multiplicity UPC trigger.   Run 3 has been proposed to include a short oxygen-oxygen run, which should offer the opportunity to study incoherent $\rho$ photoproduction on an intermediate mass target. 


 
\section*{Acknowledgements}

This work is supported in part by the U.S. Department of Energy, Office of Science, Office of Nuclear Physics, under contract number DE- AC02-05CH11231.


\begin{thebibliography}{}

\bibitem{Baltz:2007kq}
A.~J.~Baltz  \textit{et al.}
Phys. Rept. \textbf{458} (2008), 1-171
doi:10.1016/j.physrep.2007.12.001
[arXiv:0706.3356 [nucl-ex]].

\bibitem{Bertulani:2005ru}
C.~A.~Bertulani, S.~R.~Klein and J.~Nystrand,
Ann. Rev. Nucl. Part. Sci. \textbf{55} (2005), 271-310
doi:10.1146/annurev.nucl.55.090704.151526
[arXiv:nucl-ex/0502005 [nucl-ex]].

\bibitem{Klein:2020fmr}
S.~Klein and P.~Steinberg,
Ann. Rev. Nucl. Part. Sci. \textbf{70} (2020), 323-354
doi:10.1146/annurev-nucl-030320-033923
[arXiv:2005.01872 [nucl-ex]].

\bibitem{Contreras:2015dqa}
J.~G.~Contreras and J.~D.~Tapia Takaki,
Int. J. Mod. Phys. A \textbf{30} (2015), 1542012
doi:10.1142/S0217751X15420129

\bibitem{Klein:2020jom}
S.~Klein, A.~H.~Mueller, B.~W.~Xiao and F.~Yuan,
Phys. Rev. D \textbf{102} (2020) no.9, 094013
doi:10.1103/PhysRevD.102.094013
[arXiv:2003.02947 [hep-ph]].

\bibitem{STAR:2002caw}
C.~Adler \textit{et al.} [STAR],
Phys. Rev. Lett. \textbf{89} (2002), 272302
doi:10.1103/PhysRevLett.89.272302
[arXiv:nucl-ex/0206004 [nucl-ex]].

\bibitem{Bauer:1977iq}
T.~H.~Bauer, R.~D.~Spital, D.~R.~Yennie and F.~M.~Pipkin,
Rev. Mod. Phys. \textbf{50} (1978), 261
[erratum: Rev. Mod. Phys. \textbf{51} (1979), 407]
doi:10.1103/RevModPhys.50.261

\bibitem{Baur:2003ar}
G.~Baur, K.~Hencken, A.~Aste, D.~Trautmann and S.~R.~Klein,
Nucl. Phys. A \textbf{729} (2003), 787-808
doi:10.1016/j.nuclphysa.2003.09.006
[arXiv:nucl-th/0307031 [nucl-th]].

\bibitem{Baltz:2002pp}
A.~J.~Baltz, S.~R.~Klein and J.~Nystrand,
Phys. Rev. Lett. \textbf{89} (2002), 012301
doi:10.1103/PhysRevLett.89.012301
[arXiv:nucl-th/0205031 [nucl-th]].

\bibitem{Baltz:1996as}
A.~J.~Baltz, M.~J.~Rhoades-Brown and J.~Weneser,
Phys. Rev. E \textbf{54} (1996), 4233-4239
doi:10.1103/PhysRevE.54.4233

\bibitem{ALICE:2008ngc}
K.~Aamodt \textit{et al.} [ALICE],
JINST \textbf{3} (2008), S08002
doi:10.1088/1748-0221/3/08/S08002

\bibitem{ALICE:2020ugp}
S.~Acharya \textit{et al.} [ALICE],
JHEP \textbf{06} (2020), 035
doi:10.1007/JHEP06(2020)035
[arXiv:2002.10897 [nucl-ex]].

\bibitem{ALICE:2021jnv}
S.~Acharya \textit{et al.} [ALICE],
Phys. Lett. B \textbf{820} (2021), 136481
doi:10.1016/j.physletb.2021.136481
[arXiv:2101.02581 [nucl-ex]].

\bibitem{Klein:2016yzr}
S.~R.~Klein, J.~Nystrand, J.~Seger, Y.~Gorbunov and J.~Butterworth,
Comput. Phys. Commun. \textbf{212} (2017), 258-268
doi:10.1016/j.cpc.2016.10.016
[arXiv:1607.03838 [hep-ph]].

\bibitem{Gupta:1955zza}
S.~N.~Gupta,
Phys. Rev. \textbf{99} (1955), 1015-1019
doi:10.1103/PhysRev.99.1015

\bibitem{Klein:1999qj}
S.~Klein and J.~Nystrand,
Phys. Rev. C \textbf{60} (1999), 014903
doi:10.1103/PhysRevC.60.014903
[arXiv:hep-ph/9902259 [hep-ph]].

\bibitem{Guzey:2013jaa}
V.~Guzey, M.~Strikman and M.~Zhalov,
Eur. Phys. J. C \textbf{74} (2014) no.7, 2942
doi:10.1140/epjc/s10052-014-2942-z
[arXiv:1312.6486 [hep-ph]].

\bibitem{Cepila:2016uku}
J.~Cepila, J.~G.~Contreras and J.~D.~Tapia Takaki,
Phys. Lett. B \textbf{766} (2017), 186-191
doi:10.1016/j.physletb.2016.12.063
[arXiv:1608.07559 [hep-ph]].

\bibitem{Goncalves:2017wgg}
V.~P.~Gon\c{c}alves, M.~V.~T.~Machado, B.~D.~Moreira, F.~S.~Navarra and G.~S.~dos Santos,
Phys. Rev. D \textbf{96} (2017) no.9, 094027
doi:10.1103/PhysRevD.96.094027
[arXiv:1710.10070 [hep-ph]].

\bibitem{Broz:2019kpl}
M.~Broz, J.~G.~Contreras and J.~D.~Tapia Takaki,
Comput. Phys. Commun. \textbf{253} (2020), 107181
doi:10.1016/j.cpc.2020.107181
[arXiv:1908.08263 [nucl-th]].

\bibitem{Klein:2016dtn}
S.~R.~Klein [STAR],
PoS \textbf{DIS2016} (2016), 188
doi:10.22323/1.265.0188
[arXiv:1606.02754 [nucl-ex]].

\bibitem{ZEUS:2011tzw}
H.~Abramowicz \textit{et al.} [ZEUS],
Eur. Phys. J. C \textbf{72} (2012), 1869
doi:10.1140/epjc/s10052-012-1869-5
[arXiv:1111.4905 [hep-ex]].

\bibitem{Antonioli:2013ppp}
P.~Antonioli \textit{et al.} [ALICE],
CERN-LHCC-2013-019.

\bibitem{Citron:2018lsq}
Z.~Citron \textit{et al.}
CERN Yellow Rep. Monogr. \textbf{7} (2019), 1159-1410
doi:10.23731/CYRM-2019-007.1159
[arXiv:1812.06772 [hep-ph]].

\bibitem{ALICE:2021wim}
 [ALICE],
ALICE-PUBLIC-2021-004.

\bibitem{Brewer:2021kiv}
J.~Brewer, A.~Mazeliauskas and W.~van der Schee,
[arXiv:2103.01939 [hep-ph]].

\bibitem{Klein:2019qfb}
S.~R.~Klein and H.~M\"antysaari,
Nature Rev. Phys. \textbf{1} (2019) no.11, 662-674
doi:10.1038/s42254-019-0107-6
[arXiv:1910.10858 [hep-ex]].

\bibitem{IAEA}https://nds.iaea.org/relnsd/vcharthtml/VChartHTML.html

\end{thebibliography}
\end{document}